\documentclass[12pt]{iopart}
\usepackage{bm}
\usepackage{graphicx}
\usepackage{iopams}
\begin{document}
\title[Electron elastic scattering off endohedral fullerenes...]{Electron elastic scattering off endohedral fullerenes $A$@C$_{60}$: The initial insight}
\author{ V K Dolmatov,  M B Cooper and M E Hunter}
\address{Department of Physics and Earth Science, University of North Alabama,
Florence, Alabama 35632, USA}
\ead{vkdolmatov@una.edu}
\begin{abstract}
The initial insight into electron elastic scattering off endohedral
fullerenes $A$@C$_{60}$  is gained in the framework of a theoretical approach where the C$_{60}$ cage is modelled by a rectangular
(in the radial coordinate) potential well, as in many other $A$@C$_{60}$ studies.
The effect  of a noticeably weaker electron elastic scattering off $A$@C$_{60}$
compared to that off empty C$_{60}$ or even the isolated atom $A$ itself, as well as a strong sensitivity of $\rm e$ $+$ $A$@C$_{60}$ scattering
to the spin of the captured atom $A$ are unraveled,
for certain kinds of atoms. Obtained results lay out the initial
qualitative basis for identifying  interesting measurements and/or more rigorous calculations of  $\rm e$~$+$~$A$@C$_{60}$  elastic scattering to perform.

\end{abstract}
\submitto{\jpb}
\maketitle
\section{Introduction}

Spectra of $A$@C$_{60}$ endohedral fullerenes, where the atom $A$ is being encapsulated inside the hollow interior
of the C$_{60}$ cage, have attracted a great deal of attention of researchers. To date, primarily photoionization
spectra of various atoms $A$ from $A$@C$_{60}$ systems have been   detailed  theoretically at various levels of
sophistication (see, e.g., review papers \cite{RPC04,AQC09} as well as some recent works on the subject
\cite{LeePRA13,Rmatrix12,HimadriZn@C60,AmusiaJETPL09} and references therein) and,
 to a lesser extent, experimentally \cite{Exp10,Exp13}. Some insight has also been gained by theorists into
other basic phenomena of nature occurring in $A$@C$_{60}$, such as Auger vacancy decays \cite{Averbukh06,Amusia_Auger06,Solovyov11} and
fast charged-particle impact ionization reaction \cite{e+H@C60,e+Xe@C60}.
However, to the best of the authors' knowledge, another important basic phenomenon of nature, namely, electron elastic
scattering off a quantum target, has, so far, escaped   study for the case of electron scattering off $A$@C$_{60}$, despite its basic significance.

It is the ultimate aim of the
present paper to reveal possible trends in $\rm e$ $+$ $A$@C$_{60}$ elastic scattering associated with the nature of an encapsulated atom, its size
and spin. Specifically, it is discovered   that placing an atom $A$ inside the C$_{60}$ can
make electron scattering
off $A$@C$_{60}$  weaker than off the  empty C$_{60}$ cage. The effect is shown to be considerably enhanced if an encapsulated atom $A$ donates
a noticeable part of its valence electron density to the C$_{60}$ cage. Moreover, it is shown that, for such $A$@C$_{60}$ fullerenes, electron scattering
can even be weaker than off the isolated atom $A$ itself. Furthermore, it is found that if such encapsulated atom has also a nonzero spin, then
 $\rm e$ $+$ $A$@C$_{60}$ elastic scattering becomes strongly spin-dependent.

 Atomic units are used throughout the paper unless specified otherwise.

\section{Theory in brief}

Electron scattering off a multielectron target is too challenging for theorists even with regard to a free atom,
not to mention a $A$@C$_{60}$ target. In the present work, the authors opt  for
a simplified theoretical approach to e$+A@{\rm C_{60}}$ scattering. The aim is to uncover effects which \textit{might} occur in
$\rm e$ $+$ $A$@C$_{60}$  scattering rather than to perform rigorous calculations for one particular  $\rm e$ $+$ $A$@C$_{60}$ system. Thus, in the model,
(a) electron correlation is omitted from consideration, (b) both the encapsulated atom $A$ and C$_{60}$ cage are regarded as non-polarizable targets
and (c)   the C$_{60}$ cage itself is modelled by a rectangular
(in the radial coordinate) potential well $U_{\rm c}(r)$, as in many of the above cited $A$@C$_{60}$ photoionization studies:
\begin{eqnarray}
U_{\rm c}(r)=\left\{\matrix {
-U_{0}, & \mbox{if $r_{0} \le r \le r_{0}+\Delta$} \nonumber \\
0 & \mbox{otherwise.} } \right.
\label{SWP}
\end{eqnarray}
Here, the adjustable  parameters $r_{0}$ and $\Delta$ are, respectively, the inner radius and thickness of the C$_{60}$ cage and $U_{0}$ is the potential well depth.
Furthermore,  the wavefunctions $\psi_{n \ell m_{\ell} m_{s}}({\bi r}, \sigma)=r^{-1}P_{nl}(r)Y_{l m_{\ell}}(\theta, \phi) \chi_{m_{s}}(\sigma)$
and binding energies $\epsilon_{n l}$ of atomic electrons
  ($n$, $\ell$,  $m_{\ell}$ and $m_{s}$ is the standard set of quantum numbers of an electron in a central field, $\sigma$ is the electron spin coordinate) will be the solutions of `endohedral' Hartree-Fock (HF) equations:
\begin{eqnarray}
&&\left[ -\frac{\Delta}{2} - \frac{Z}{r} +U_{\rm c}(r) \right]\psi_{i}
({\bi x}) + \sum_{j=1}^{Z} \int{\frac{\psi^{*}_{j}({\bi x'})}{|{\bi
x}-{\bi x'}|}} \nonumber \\
 && \times[\psi_{j}({\bi x'})\psi_{i}({\bi x})
- \psi_{i}({\bi x'})\psi_{j}({\bi x})]d {\bi x'} =
\epsilon_{i}\psi_{i}({\bi x}). \label{eqHF}
\end{eqnarray}
Here, $Z$ is the nuclear charge of the atom, ${\bi x}=({\bi r}, \sigma)$ and the integration over ${\bi x}$ implies both the integration over ${\bi r}$ and summation over
$\sigma$.
The `endohedral' HF equation, obviously, differs from the ordinary HF equation for a free atom by the presence of the $U_{\rm c}(r)$ potential in the former. To solve the problem of ${\rm e}+A@{\rm C_{60}}$ scattering,
one first solves  \eref{eqHF}  for the ground state of the atom, thereby calculating the  wavefunctions of the atomic electrons. Once  all atomic functions are determined, they are plugged back  into \eref{eqHF} to calculate scattering state
 wavefunctions [now being $\psi_{i}({\bi x})$ functions in \eref{eqHF}] and, thus, their radial parts $P_{\epsilon_{i}\ell_{i}}(r)$. Corresponding electron elastic scattering phase shift $\delta_{\ell}(\epsilon)$ is then determined by
 referring to the asymptotic behaviour of $P_{\epsilon\ell}(r)$ at $r\gg 1$:
\begin{eqnarray}
P_{\epsilon\ell}(r) \approx \frac{1}{\sqrt{\pi k}}\sin\left(k r -\frac{\pi\ell}{2}+\delta_{\ell}(\epsilon)\right),
\label{P(r)}
\end{eqnarray}
with $k = \sqrt{2\epsilon}$ being the momentum of a scattered electron. The total electron elastic scattering cross sections $\sigma(\epsilon)$ is then calculated as follows:
 \begin{eqnarray}
 \sigma(\epsilon)= \frac{4\pi}{k^2}\sum^{\infty}_{\ell=0}(2\ell+1)\sin^{2}\delta_{\ell}(\epsilon).
 \label{sigma}
 \end{eqnarray}

One of the atoms of  concern of this study is the Mn(...$\rm 3d^{5}$$\rm 4s^{2}$, $^{6}$S) atom which has the total spin of $5/2$ owing to the presence of a ${\rm 3d}^{5}$ semifilled subshell in the atomic
ground-state configuration. Such atoms require a separate theoretical treatment. A convenient, effective
  theory to calculate the structure of a semifilled shell atom  is a `spin-polarized' Hartree-Fock (SPHF) approximation developed by Slater \cite{Slater}.  The quintessence of SPHF is as follows. It accounts for the fact that spins of all
electrons in a semifilled subshell of the atom (e.g., in the $\rm 3d^{5}$ subshell of Mn) are co-directed, in accordance with Hund's rule, say, all pointing upward. This results in splitting of each of other
closed ${n\ell}^{2(2\ell+1)}$ subshells in the atom into two semifilled subshells of opposite spin orientations, ${n\ell}^{2\ell+1}$$\uparrow$ and ${n\ell}^{2\ell+1}$$\downarrow$. This is in view of
 the presence of
exchange interaction between $nl$$\uparrow$ electrons with  spin-up electrons in the original semifilled
subshell of the atom (like the $\rm 3d^{5}$$\uparrow$ subshell in the Mn atom) but  absence of such for $nl$$\downarrow$ electrons. Thus, the Mn atom has the following SPHF configuration:
 Mn(...${3\rm p}^{3}$$\uparrow$${3\rm p}^{3}$$\downarrow$${3\rm d}^{5}$$\uparrow$${4\rm s}^{1}$$\uparrow$$4s^{1}$$\downarrow$, $^{6}$S).
SPHF equations for the ground, bound excited and scattering states of a semifilled shell atom differ from ordinary HF equations for closed shell atoms by accounting for exchange interaction only between electrons with the same spin orientation
($\uparrow$, $\uparrow$ or $\downarrow$, $\downarrow$). To date, SPHF  has successfully been extended to studies of electron elastic scattering off isolated semifilled shell atoms in a number of works \cite{JETP,PRAe+Mn,Remeta} (and references therein). In the present paper, SPHF is utilized for calculation of both the structure of ${\rm Mn@C_{60}}$  and $\rm e$ $+$ Mn@C$_{60}$ scattering. This is achieved on the
basis of the `endohedral' HF  equations (\ref{eqHF}) where exchange interaction is now accounted for only between  electrons   with the same spin direction.

In conclusion, previously, Winstead and McKoy \cite{McKoy06} studied electron elastic scattering off \textit{empty} C$_{60}$ by modelling the C$_{60}$ cage by the same square-well potential \eref{SWP}.
 In addition, they investigated it in the framework of a much more
elaborated \textit {ab initio} `static-exchange' approximation as well \cite{McKoy06}.  In the static-exchange approximation for $\rm e$ $+$ C$_{60}$  scattering, the electron density of C$_{60}$ is regarded the same as that in the C$_{60}$ ground-state throughout a scattering process, i.e., is `frozen' or `static'. It is calculated in the framework of an \textit{ab initio} Hartree-Fock approximation applied to the C$_{60}$ molecule. After that, the Schwinger multichannel theory is utilized along with accounting for distinct irreducible representations of the $I_{\rm h}$ point group for partial electronic waves of scattered electrons in order to complete the
$\rm e$ $+$ C$_{60}$ scattering study.
 A  detailed comparison of calculated results obtained in the semi-empirical square-well and {\textit ab initio} static-exchange approximations, performed by Winstead and McKoy \cite{McKoy06},
 revealed a qualitative, and even semiquantitative, agreement between some of the most prominent features of $\rm e$ $+$ C$_{60}$ elastic scattering predicted by the square-well and far more superior `static-exchange' approximation. Interesting,  even a yet greater simplified
  model, where the C$_{60}$ cage  was simulated by an infinitesimally thin $\delta$-function-like potential \cite{e+C60}, predicts some of the above mentioned major features as well \cite{McKoy06}; the predictions,
  however, are noticeably different with respect to their strengths and positions compared to results of the two other calculations.  The discussed results of work \cite{McKoy06} on $\rm e$ $+$ C$_{60}$
  scattering give us confidence in that
the approach to electron elastic scattering off \textit{endohedral fullerenes}, utilized in the present paper, is usable for getting
the initial insight into $\rm e$ $+$ $A$@C$_{60}$ elastic scattering. The gained insight, in turn,
 will identify future interesting measurements and/or detailed  calculations to perform.

\section{Results and discussion}

In  performed calculations, the authors employ the same values of $r_{0}$, $\Delta$ and $U_{0}$ as in the work by Winstead and McKoy \cite{McKoy06} for $\rm e$ $+$ C$_{60}$  elastic scattering. Namely,
$\Delta = 2.9102$ (which is twice of the covalent radius of carbon), $r_{0} = 5.262 = R_{\rm c} - 1/2\Delta$ ($R_{\rm c}=6.7173$ being the radius of the C$_{60}$ skeleton)
and $U_{0} = 7.0725$ eV (which was found  by matching the electron affinity $EA=-2.65$ eV of C$_{60}$ with the assumption that the orbital momentum of the $2.65$-eV-state is $\ell =1$ \cite{McKoy06}). The  choice is dictated by that the given
 values of the square-well potential parameters were shown  \cite{McKoy06} to lead to a qualitative and even semiquantitative agreement between both the square-well model and `static-exchange' approximation calculated data
 for $\rm e$ $+$~ C$_{60}$  scattering.
In all  present calculations, ten partial electronic waves with the orbital momentum $\ell$ up to $\ell_{max}=9$ are accounted for in the $\ell$-summation in  \eref{sigma}. The contributions
of terms with $\ell > 9$ in (\ref{sigma}) were found to be insignificant in the presently considered electron energy domain of $\epsilon \leq 15$ eV, where the most interesting phenomena occur.

\subsection{Total electron elastic scattering cross sections $\sigma_{\rm el}^{\uparrow}$ and $\sigma_{\rm el}^{\downarrow}$ of ${\rm Mn@C_{60}}$}

\begin{figure}[h]
\center{\includegraphics[width=8cm]{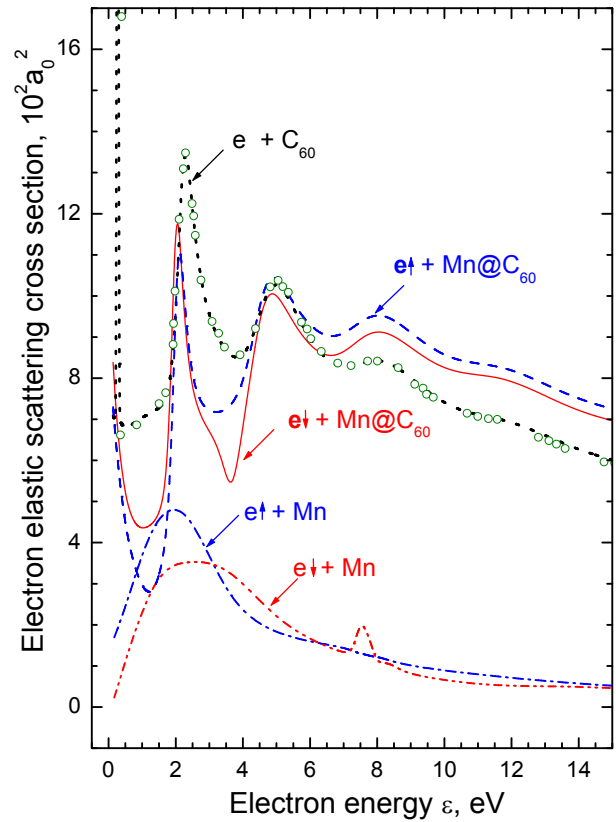}}
\caption{Calculated electron elastic scattering cross sections $\sigma_{\rm el}(\epsilon)$ (in units of $10^{2}a_{0}^2$, $a_{0}$ being the first Bohr radius) for spin-up and spin-down electron scattering off Mn@C$_{60}$ and C$_{60}$, obtained
in the framework of the square-well potential model with $r_{0} = 5.262$, $\Delta = 2.9102$ and $U_{0} = 7.0725$ eV (dotted line, present calculations; open circles, data from \cite{McKoy06}) as well as off free Mn \cite{PRAe+Mn}
(the latter are results of a
sophisticated RPAE calculation which accounted for polarization and correlation effects in $\rm e$$\uparrow$$\downarrow$ $+$ Mn scattering).
}
\label{fig1}
\end{figure}

An amazing example of the `all-in-one' case study, where the novel findings of the performed research are all present at the same time, is  electron elastic scattering off Mn@C$_{60}$. Corresponding calculated
data both for the spin-up ($\rm e$$\uparrow$ $+$ Mn@C$_{60}$) and spin-down ($\rm e$$\downarrow$ $+$ Mn@C$_{60}$) electron elastic scattering cross sections
$\sigma_{\rm el}^{\uparrow (\downarrow)}(\epsilon)$ are depicted in \fref{fig1} along with present calculated data for scattering just off empty C$_{60}$ as well as off free Mn \cite{PRAe+Mn}.

First, note that the encapsulation of the Mn atom in the hollow inner space of C$_{60}$ makes electron scattering off ${\rm Mn@C_{60}}$ to differ greatly from scattering off the same sized empty C$_{60}$ cage.
It is, thus, shown that the C$_{60}$ cage cannot `hide' the presence of the encapsulated atom inside the cage from the `attention' of incoming electrons. The implication is that electron scattering off fullerenes can,
in principle, be
controlled by encapsulating atoms inside the C$_{60}$ cage.

Second, amazingly, note how electron scattering of ${\rm Mn@C_{60}}$ appears to be much weaker than electron scattering off empty C$_{60}$ in ceratin electron energy domains, first between   $\epsilon \approx 0$
and $1.9$ eV, then, once again, between $\epsilon \approx 2.2$ and $4.2$ eV. This finding implies that a gas phase medium of endohedral fullerenes $A$@C$_{60}$ can, in principle, be less resistive to the propagating
incoming electrons than a gaseous medium of the same sized empty fullerenes C$_{60}$.

Third, even more amazingly, note how the electron scattering cross section of ${\rm Mn@C_{60}}$ is smaller than that of a free Mn atom itself, in the electron energy region between  $\epsilon \approx 0.8$
and $1.7$ eV. The implication is that a gas phase medium of endohedral fullerenes $A$@C$_{60}$ can, in principle, be less resistive to the incoming electrons than  a gaseous medium of smaller sized free atoms.

Fourth, note how generally different from each other are  the electron spin-up $\sigma_{\rm el}^{\uparrow}(\epsilon)$ and spin-down $\sigma_{\rm el}^{\downarrow}(\epsilon)$ scattering cross
sections, especially at lower electron energies. Thus, not only electron elastic scattering off $A$@C$_{60}$ can reveal the presence of the atom inside C$_{60}$, but it can determine the existence of a non-zero spin of the captured
atom as well. Electron scattering off $A$@C$_{60}$, where the atom $A$ has a non-zero spin, can, thus, in principle, serve as a tool for producing outgoing spin-polarized electron beams in gaseous media of
endohedral fullerenes.

\subsection{Partial electron elastic scattering cross sections  $\sigma_{\ell}^{\uparrow}$ and $\sigma_{\ell}^{\downarrow}$ of ${\rm Mn@C_{60}}$}

\begin{figure}[h]
\center{\includegraphics[width=8cm]{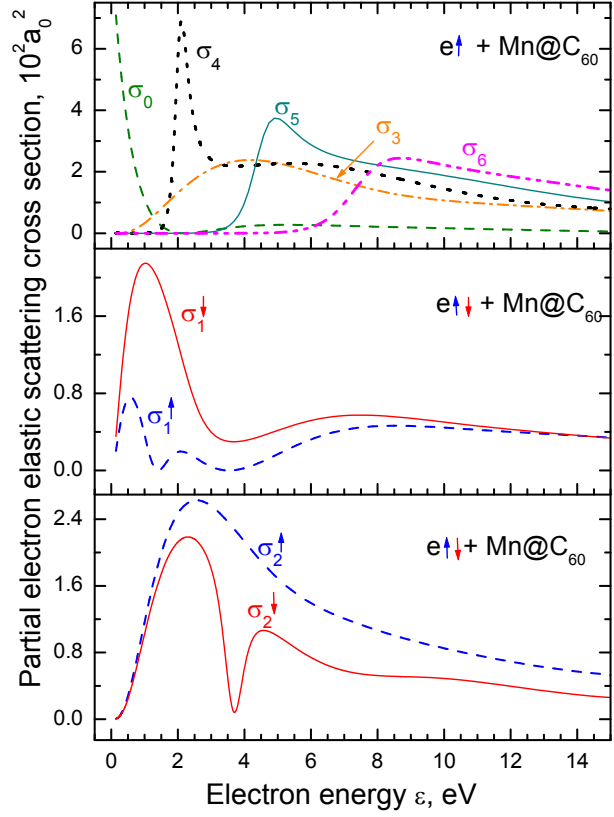}}
\caption{Calculated partial electron elastic scattering cross sections $\sigma_{\ell}^{\uparrow(\downarrow)}(\epsilon)$ of Mn@C$_{60}$, as marked. For $\ell \ge 3$,
corresponding spin-up and spin-down cross sections are practically the same; therefore, only $\sigma_{\ell \ge 3}^{\uparrow}$ are plotted.}
\label{fig2}
\end{figure}

In order to better understand the $\rm e$$\uparrow$($\downarrow$) $+$ Mn@C$_{60}$ total electron elastic scattering cross sections,  depicted in \fref{fig1}, let us explore corresponding partial electron elastic scattering cross sections
$\sigma_{\ell}^{\uparrow}$ and $\sigma_{\ell}^{\downarrow}$  one by one for different $\ell$s, \fref{fig2}.

One can see that each of $\sigma_{\ell}^{\uparrow(\downarrow)}$ has at least one major maximum. By comparing  figures \ref{fig1} and \ref{fig2}, one arrives at the conclusion that
the maxima in $\sigma_{\rm el}^{\uparrow(\downarrow)}$ at $\epsilon \approx 8.5$,  $5$ and $2$  eV are primarily due to corresponding major maxima in the partial cross sections with $\ell=6$, $5$ and $4$, respectively.
The trial calculations showed that the same remains true for the origin of maxima seen in the electron elastic scattering cross section off empty C$_{60}$ at the same energies $\epsilon \approx 8.5$,  $5$ and $2$  eV
(\fref{fig1}). The latter is in line with the conclusions of work \cite{McKoy06} where the  $\rm e$ $+$ C$_{60}$ scattering cross section was studied previously. Furthermore, by exploring figures \ref{fig1} and \ref{fig2}
one can understand where the major differences between the $\sigma_{\rm el}^{\uparrow}$ and $\sigma_{\rm el}^{\downarrow}$ total cross
sections of the Mn@C$_{60}$ system come from. Clearly, they are brought about primarily by the differences between the $\sigma_{\ell=2}^{\uparrow}$ and $\sigma_{\ell=2}^{\downarrow}$ partial cross sections,
on the one hand, and   $\sigma_{\ell=1}^{\uparrow}$ and $\sigma_{\ell=1}^{\downarrow}$, on the other hand.

\subsection{$\rm e$$\uparrow$$\downarrow$ $+$ Mn@C$_{60}$ elastic scattering phase shifts $\delta_{\ell }^{\uparrow(\downarrow)}(\epsilon)$}

\begin{figure}[h]
\center{\includegraphics[width=8cm]{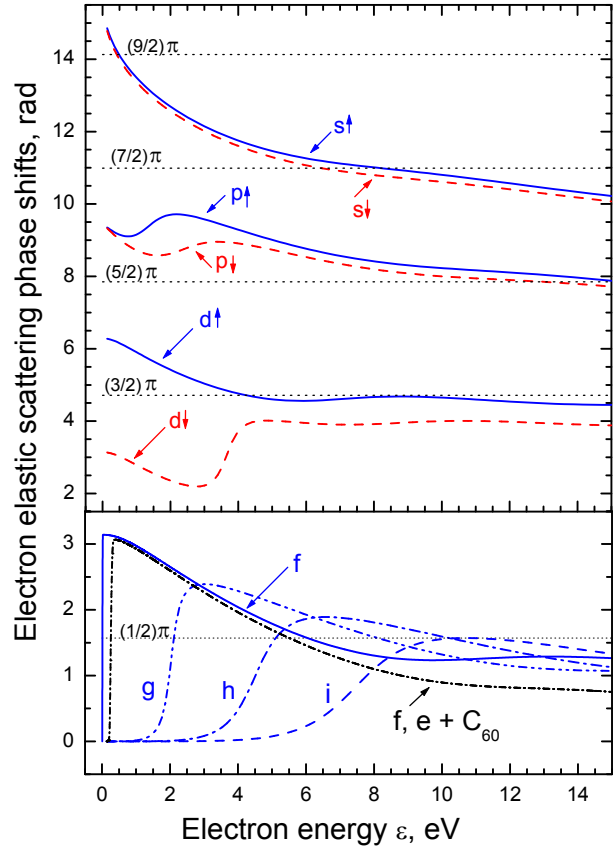}}
\caption{Calculated electron elastic scattering phase shifts $\delta_{\ell}^{\uparrow(\downarrow)}$ for $\rm e$$\uparrow$$\downarrow$ $+$ Mn@C$_{60}$ scattering.
For electrons with $\ell \ge 3$, spin-up and spin-down phase shifts are practically the same; therefore, only $\delta_{\ell \ge 3}^{\uparrow}$  are plotted.}
\label{fig3}
\end{figure}

In order to get insight into the behaviour of the partial $\rm e$$\uparrow$$\downarrow$ $+$ Mn@C$_{60}$  electron elastic scattering cross sections $\sigma_{\ell}^{\uparrow(\downarrow)}$ and, thus,
 better understand the $\rm e$$\uparrow$$\downarrow$ $+$ Mn@C$_{60}$
scattering reaction itself, let us explore corresponding partial
electron elastic scattering phase shifts $\delta_{\ell}^{\uparrow\downarrow}(\epsilon)$, \fref{fig3}.

First, note the values which the $\delta_{\ell}^{\uparrow}$ and $\delta_{\ell}^{\downarrow}$ phase shifts take at $\epsilon = 0$. They are consistent with the well known
Levinson's theorem of scattering theory. In our case, the latter can be stated as $\delta_{\ell }^{\uparrow(\downarrow)}(0) = (n_{\ell}^{\uparrow(\downarrow)}+q_{\ell }^{\uparrow(\downarrow)})\pi $. Here,
 $n_{\ell }^{\uparrow}$  ($n_{\ell }^{\downarrow}$) is the number
of bound states with given $\ell^{\uparrow}$ ($\ell^{\downarrow}$) in the field of a $A$@C$_{60}$ system, whereas  $q_{\ell }^{\uparrow}$ and $q_{\ell }^{\downarrow}$
are the number of occupied $\ell^{\uparrow}$ and $\ell^{\downarrow}$ states, respectively, in the encapsulated atom itself. A trial calculation showed that
 only extra  $\rm s$$\uparrow$$\downarrow$-, $\rm p$$\uparrow$$\downarrow$- and $\rm d$$\uparrow$$\downarrow$-bound states emerge in the field of Mn@C$_{60}$ in addition to those already existing in the Mn atom. Correspondingly,
as in \fref{fig3}, $\delta_{\ell=2}^{\uparrow}(0) = 2\pi$ but  $\delta_{\ell=2}^{\downarrow}(0) = \pi$, because of the presence of a  $\rm 3d^{5}$$\uparrow$ subshell but the absence of a $\rm 3d^{5}$$\downarrow$ subshell
 in the Mn atom configuration. Other phase shifts with given $\ell$ but opposite electron spin orientations have, respectively, the same values at $\epsilon =0$:
 $\delta_{\ell=0}^{\uparrow(\downarrow)}(0) = 5\pi$, $\delta_{\ell=1}^{\uparrow(\downarrow)}(0) = 3\pi$ and  $\delta_{\ell \ge 3}^{\uparrow(\downarrow)}(0) = 0$, as in \fref{fig3}.

Second, by comparing figures \ref{fig3} and \ref{fig2} one arrives at the understanding that the maxima in all Mn@C$_{60}$ partial scattering cross sections $\sigma_{\ell}^{\uparrow(\downarrow)}$  occur where corresponding phase shifts pass through, or close to, the values of $\delta_{\ell}^{\uparrow(\downarrow)} = n\pi/2$ where $\sin{\delta_{\ell}}$ reaches its maximum value.

\subsection{Reasons for the marked differences between $e$ $+$ C$_{60}$ and $e$ $+$ Mn@C$_{60}$ scattering}

\begin{figure}[h]
\center{\includegraphics[width=8cm]{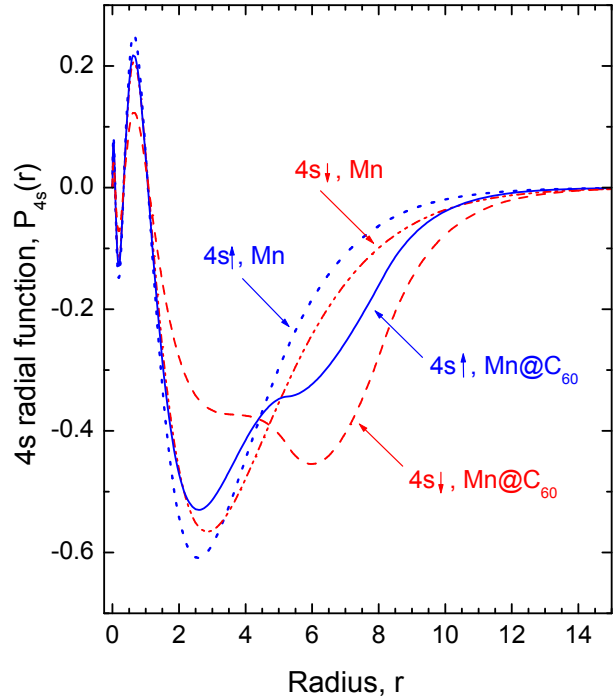}}
\caption{Calculated $P_{\rm 4s\uparrow}(r)$ and $P_{\rm 4s\downarrow}(r)$ radial functions of the $\rm 4s$$\uparrow$ and $\rm 4s$$\downarrow$ electrons of Mn@C$_{60}$ and free Mn.}
\label{fig4}
\end{figure}

Depicted in \fref{fig4} are calculated data for the $P_{\rm 4s\uparrow}(r)$ and $P_{\rm 4s\downarrow}(r)$ radial functions of the valence $\rm 4s$$\uparrow$ and $\rm 4s$$\downarrow$ electrons of free Mn and Mn@C$_{60}$.

One can see that the encapsulation of the Mn atom inside the C$_{60}$ cage results in that the $P_{\rm 4s\uparrow}(r)$ and $P_{\rm 4s\downarrow}(r)$ functions of the atom become noticeably drawn into the region
of the square-well potential. This implies a noticeable electron density transfer from the encapsulated atom to the C$_{60}$ cage. Therefore, electron scattering off Mn@C$_{60}$
occurs in a  different potential than in the case of scattering off empty C$_{60}$. The difference between these potentials seems to be strong enough to make electron elastic scattering off Mn@C$_{60}$
to be greatly different than scattering off empty C$_{60}$. One can generally state that if a captured atom $A$ in $A$@C$_{60}$ transfers a significant portion of its electron density to the C$_{60}$ cage, then electron scattering off
$A$@C$_{60}$ should differ markedly from electron scattering off empty C$_{60}$. This statement will find
 a supporting evidence later in the paper where electron elastic scattering off three other targets, namely, Ar@C$_{60}$, Xe@C$_{60}$ and Ba@C$_{60}$ is discussed.

\subsection{Reasons for a marked electron spin dependence of $e$ $+$ Mn@C$_{60}$ scattering}

Another novel finding here is that the $\rm 4s$$\uparrow$ and
$\rm 4s$$\downarrow$ electrons of Mn appear to donate  different amounts of their electron densities to the C$_{60}$ cage, see \fref{fig4}. Namely, the $\rm 4s$$\downarrow$ electron donates noticeably more of its electron density than the  $\rm 4s$$\uparrow$  electron. This, results, figuratively  speaking, in `charging' the C$_{60}$ cage with a
\textit{spin-down} electron  density by a  \textit{spin-neutral} $\rm 4s^{2}$ subshell.  In the present paper, the discovered effect is referred to as
the `C$_{60}$-spin-charging effect'.

The `C$_{60}$-spin-charging' effect sheds more light on a reason as to why electron elastic scattering off Mn@C$_{60}$ is strongly spin-dependent.
Because, in Mn@C$_{60}$, the C$_{60}$ cage becomes overall `spin-down-charged', exchange interaction between the incoming spin-up electrons with the C$_{60}$ cage differs from exchange interaction between
the incoming spin-down electrons and the `spin-down-charged' C$_{60}$ cage. This is in addition to differences between exchange interaction of these electrons with the semifilled $\rm 3d^{5}$$\uparrow$ subshell of
the Mn atom. As a result, the incoming spin-up electrons feel a different potential of Mn@C$_{60}$ than the incoming spin-down electrons. The differences between the two potentials appear to be strong enough to induce
marked discrepancies in scattering of spin-up and spin-down electrons off Mn@C$_{60}$. Note, the significance of the `C$_{60}$-spin-charging' effect vanishes for the incoming electrons with large orbital momenta $\ell$.
  Indeed, the greater given $\ell$  of the incoming electron, the farther it is away from the
scattering center. Exchange interaction, in turn, decreases with increasing distance between electrons. Therefore, any significant discrepancies between  scattering of spin-up and spin-down electrons off Mn@C$_{60}$ must
 vanish
for large-$\ell$-electrons. This explains why scattering of electrons with $\ell \ge 3$ off Mn@C$_{60}$ is practically spin-independent but scattering of electrons with smaller $\ell$s is not,
see figures \ref{fig2} and \ref{fig3}.

\subsection{Electron elastic scattering off Ba@C$_{60}$, Ar@C$_{60}$  and Xe@C$_{60}$}

\begin{figure}[h]
\center{\includegraphics[width=8cm]{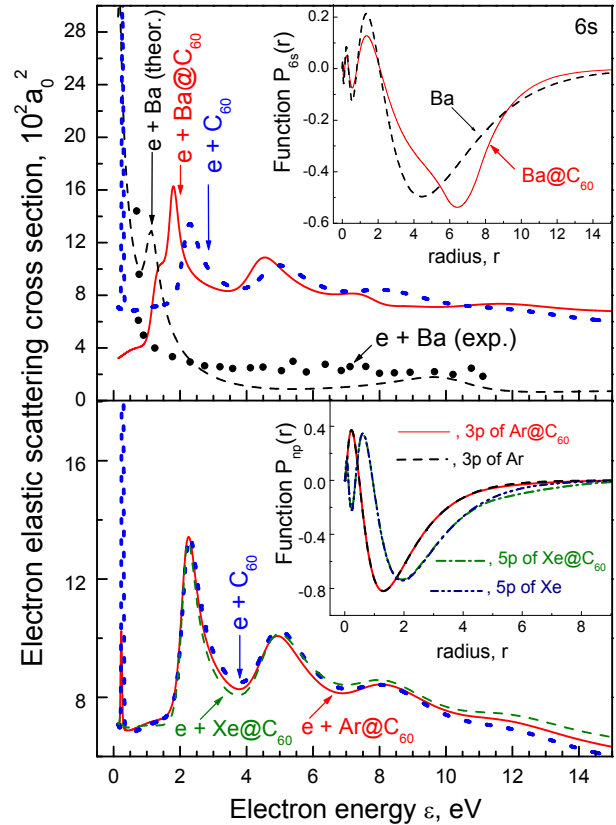}}
\caption{Upper panel: Calculated data for the electron elastic scattering cross sections $\sigma_{\rm el}(\epsilon)$ of Ba@C$_{60}$, empty C$_{60}$ and free Ba. Results labelled as `$\rm e + Ba$ (theor.)' are  present HF calculated data for $\rm e$ $+$ Ba scattering. Results labelled as `$\rm e +Ba$ (exp.)'  are corresponding experimental data  found in \cite{e+Ba}. The inset: the $P_{\rm 6s}$(r) radial functions of Ba@C$_{60}$ and free Ba. Lower panel: Calculated data for the electron elastic scattering cross sections $\sigma_{\rm el}(\epsilon)$ of Ar@C$_{60}$, Xe@C$_{60}$ and empty C$_{60}$, as marked. The inset: the $P_{\rm 3p}$(r) radial functions of Ar@C$_{60}$ and free Ar, as well as the $P_{\rm 5p}$(r) radial functions of Xe@C$_{60}$ and free Xe, as marked.}
\label{fig5}
\end{figure}

In order to verify that the above revealed trends in $\rm e$ $+$ $A$@C$_{60}$ scattering are not characteristic features of specifically Mn@C$_{60}$ but are a general occurrence, the authors calculated electron elastic
scattering off three other targets: Ba@C$_{60}$, Ar@C$_{60}$  and Xe@C$_{60}$.
Calculated data for corresponding electron elastic scattering cross sections $\sigma_{\rm el}(\epsilon)$ are depicted in \fref{fig5}.

From the upper panel of \fref{fig5}, one can see that, as in the case of Mn@C$_{60}$, the Ba@C$_{60}$ cross section $\sigma_{\rm el}(\epsilon)$ differs markedly from that of empty C$_{60}$. Also, similar to Mn@C$_{60}$,
there are energy domains where $\sigma_{\rm el}(\epsilon)$ of Ba@C$_{60}$ is noticeably smaller than $\sigma_{\rm el}(\epsilon)$ of empty C$_{60}$ or even smaller than $\sigma_{\rm el}(\epsilon)$ of free Ba.
Furthermore, note how the encapsulated Ba atom, similar to Mn, transfers a noticeable amount of its $\rm 6s^{2}$ valence electron density to the C$_{60}$ cage. One may conclude that it is primarily due to the
valence electron density transfer from the encapsulated atom  to the C$_{60}$ cage that electron scattering off a `stuffed' C$_{60}$ cage differs strongly from scattering off empty C$_{60}$.

On the other hand, one can see, in the lower panel of \fref{fig5}, that there are only insignificant discrepancies between $\sigma_{\rm el}(\epsilon)$s of Ar@C$_{60}$ and Xe@C$_{60}$, despite
Xe being a larger sized atom than Ar. Furthermore, both $\sigma_{\rm el}(\epsilon)$ of Ar and $\sigma_{\rm el}(\epsilon)$ of Xe differ little from that of empty C$_{60}$. Next, note how the encapsulated
Ar and Xe atoms both transfer little of their valence electron density  to the C$_{60}$ cage. One may conclude that where there is little transfer of the electron density from the encapsulated atom  to the C$_{60}$ cage
electron scattering off $A$@C$_{60}$ depends little on the encapsulated atom itself, regardless of its size. In such case, of course, no significant differences  emerge between electron elastic scattering off a `stuffed' and empty C$_{60}$. Both of the importance and interest, however, is that
 $\sigma_{\rm el}(\epsilon)$ of a `stuffed' C$_{60}$ (Ar@C$_{60}$ and Xe@C$_{60}$) is seen to be, once again, somewhat smaller than $\sigma_{\rm el}(\epsilon)$ of empty C$_{60}$, at certain energies.

\section{Conclusion}

The present work has provided the initial insight into possible trends in electron elastic scattering off endohedral fullerenes $A$@C$_{60}$. The most remarkable of them are (a) weaker electron elastic scattering off
$A$@C$_{60}$ than   off  empty C$_{60}$ fullerene or off a smaller sized isolated atom $A$ itself, on certain occasions and for certain atoms,  (b) a noticeable electron spin
dependence of
$\rm e$ $+$~$A$@C$_{60}$ scattering, for encapsulated semifilled shell atoms possessing a large non-zero spin, and (c) the `C$_{60}$-spin-charging' effect.  Calculated results presented in this paper bear a qualitative and, possibly, semiquantitative significance. They, however, most likely represent some of the most intrinsic properties of $\rm e$ $+$ $A$@C$_{60}$
elastic scattering, similar to the significance of the square-well model calculated results of $\rm e$~$+$ C$_{60}$ scattering \cite{McKoy06}. This is because the made predictions are independent of any fine details of bonds between the $60$ carbon atoms which make the C$_{60}$ cage. The authors hope that the present work will prompt experimentallists and other theorists to undertake corresponding studies.
Moreover, in order to understand the effects of electron correlation, target polarization and molecular structure in a $\rm e$ $+$ $A$@C$_{60}$ electron scattering process
one must know how the latter develops without accounting for these higher-order effects in calculations. The present work provides interested researchers with
such knowledge.
 Thus, results of this work will  come handy in future studies of electron elastic scattering off endohedral fullerens, thereby additionally contributing to the advancement of the field.

\ack
This work was supported by NSF Grant no. PHY-1305085.

\section*{References}

\end{document}